\documentclass[twocolumn
]{article}

\usepackage{graphicx}
\usepackage{times}

\def\apj{ApJ} 
 
\def\aap{A\&A}
\def\aaps{A\&A Supplement}

\def\solphys{Solar Phys.}

\newcommand{\ion}[2]{#1#2}
\newcommand{\acknowledgements}{\paragraph{Acknowledgements:}}
\usepackage{natbib}
\bibpunct{(}{)}{;}{a}{}{}

\begin{document}

\title{The role of emerging bipoles in the formation of a sunspot penumbra}

\author{R. Schlichenmaier%\thanks{
%  \email{schliche, nbello, rrezaei, waldmann@kis.uni-freiburg.de}\newline}%
, N. Bello Gonz\'alez, R. Rezaei, \&  T.~A. Waldmann \\ Accepted by Astronomische Nachrichten}

\maketitle

\begin{abstract}
{The generation of magnetic flux in the solar interior and its transport from the convection zone into the photosphere, the chromosphere, and the corona will be in the focus of solar physics research for the next decades. With 4\,m class telescopes, one plans to measure essential processes of radiative magneto-hydrodynamics that are needed to understand the nature of solar magnetic fields.
One key-ingredient to understand the behavior of solar magnetic field is the process of flux emergence into the solar photosphere, and how the magnetic flux reorganizes to form the magnetic phenomena of active regions like sunspots and pores. Here, we present a spectropolarimetric and imaging data set from a region of emerging magnetic flux, in which a proto-spot without penumbra forms a penumbra. During the formation of the penumbra the area and the magnetic flux of the spot increases. First results of our data analysis demonstrate that the additional magnetic flux, which contributes to the increasing area of the penumbra, is supplied by the region of emerging magnetic flux. We observe emerging bipoles that are aligned such that the spot polarity is closer to the spot. As an emerging bipole separates, the pole of the spot polarity migrates towards the spot, and finally merges with it. We speculate that this is a fundamental process, which makes the sunspot accumulate magnetic flux. As more and more flux is accumulated a penumbra forms and transforms the proto-spot into a fully-fledged sunspot.}
\end{abstract}

\section{Introduction}

The generation of the solar magnetic flux is thought to take place in the solar interior, either in the convection zone pro\-per or in the tachocline beneath it. The generated magnetic flux becomes buoyant and rises upward through the convection zone \citep{parker1979book}. Regions of flux emergence appear in the photosphere, and processes of structure formation take place. As a product pores and sunspots are  observed in the photosphere. One particularly interesting process is the formation of a sun\-spot penumbra. What is the crucial ingredient that initiates the transformation from a proto-spot into a sun\-spot with a full-size penumbra? Where does the magnetic flux come from that is needed to form the penumbra?

Our knowledge on how the penumbra forms is rather poor. \citet{zwaan1992} describes the formation of a sunspot with penumbra as coalescence of the existing pores \citep[refering to][]{mcintosh1981, bumba+suda1984}. The spatial resolution and polarimetric sensitivity of those observations did not allow to trace the smaller magnetic patches, e.g., magnetic knots or bright points. Zwaan compiles the following results: the penumbra forms section by section; each section completes within an hour; and the penumbra nearly closes leaving a gap toward the inside of the active region. 

The formation of sunspots is attended by magnetic flux emergence. In this respect, elongated granules are essential since they are signatures of flux emergence \citep{bray+loughhead1964}. This link has recently also been demonstrated in MHD simulations \citep[e.g.,][]{tortosa+moreno2009, cheung+etal2008}. Such aligned and elongated granules are compatible with the top of the magnetic loop passing through the photosphere \citep{strous+etal1996}. 

\begin{figure*}%\sidecaption
\includegraphics[width=13cm]{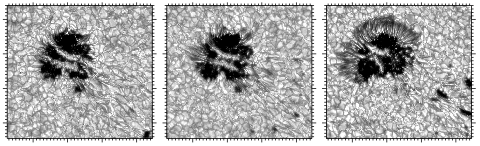}
\hfill
\parbox[b]{3.7cm}{
\caption{Snapshot images in the G-band document the evolution of a sunspot penumbra at 08:32, 09:33, and 13:06 UT. Tickmarks are in arcsec.}}
\label{fig:1}
\end{figure*}

We have recently presented a unique data set about the formation of a penumbra \citep{schlichenmaier+al2010a}. There we have established that the total area of the spot increases, and that this increase is exclusively due to an area increase of the penumbra (see also Fig.~\ref{fig:1} discussed below). This indicates that the magnetic flux of the spot is increasing, and the immediate question is: Where does the magnetic flux come from?
In this paper we investigate how the site of emerging magnetic flux is linked to the growth of the sunspot. 

\section{Observations and post-processing}\label{sec:obs}

On July 4th 2009, observations on the active region NOAA\,11024 were taken at the German Vacuum Tower Telescope in Tenerife with a multi-wavelength setup,  consisting of (1) the 2D GFPI\footnote{The GFPI system was formerly called "G\"ottingen" Fabry-P\'erot Interferometer. It was upgraded and moved to GREGOR and is now called "GREGOR" FPI.} spectro-polarimeter \citep{puschmann+etal2006, bello+kneer2008} scanning 31 spectral points along the \ion{Fe}{i}\,617.3\,nm line ($g_{\rm{eff}}$\,=\,2.5) with a step, $\Delta\lambda$\,=\,1.48\,pm, and 20\,ms exposure time; (2) the Tenerife Infrared Polarimeter \citep[TIP\,II,][]{collados+etal2007, schlichenmaier+collados2002} in the \ion{Fe}{i}\,1089.6\,nm, and (3) imaging channels in the G-band and in \ion{Ca}{ii}{\,K}. The observations were supported by the Kiepenheuer Adaptive Optics System \citep[KAOS,][]{vdluhe+etal2003}. The imaging data were reconstructed with the KISIP code \citep{woeger+ovdluhe2008}. The GFPI data were reconstructed using the "G\"ottingen'' speckle code \citep{deBoer1996}.

We construct longitudinal magnetograms applying the centre-of-gravity (COG) method \citep{rees+semel1979} to the $I\,\pm\,V$ profiles. This method provides an average of the magnetic field strength over the angular resolution element and over the formation height. The reliability of this method was discussed, e.g., in \citet{bello2006}.

\begin{figure*}
\includegraphics[width=16cm]{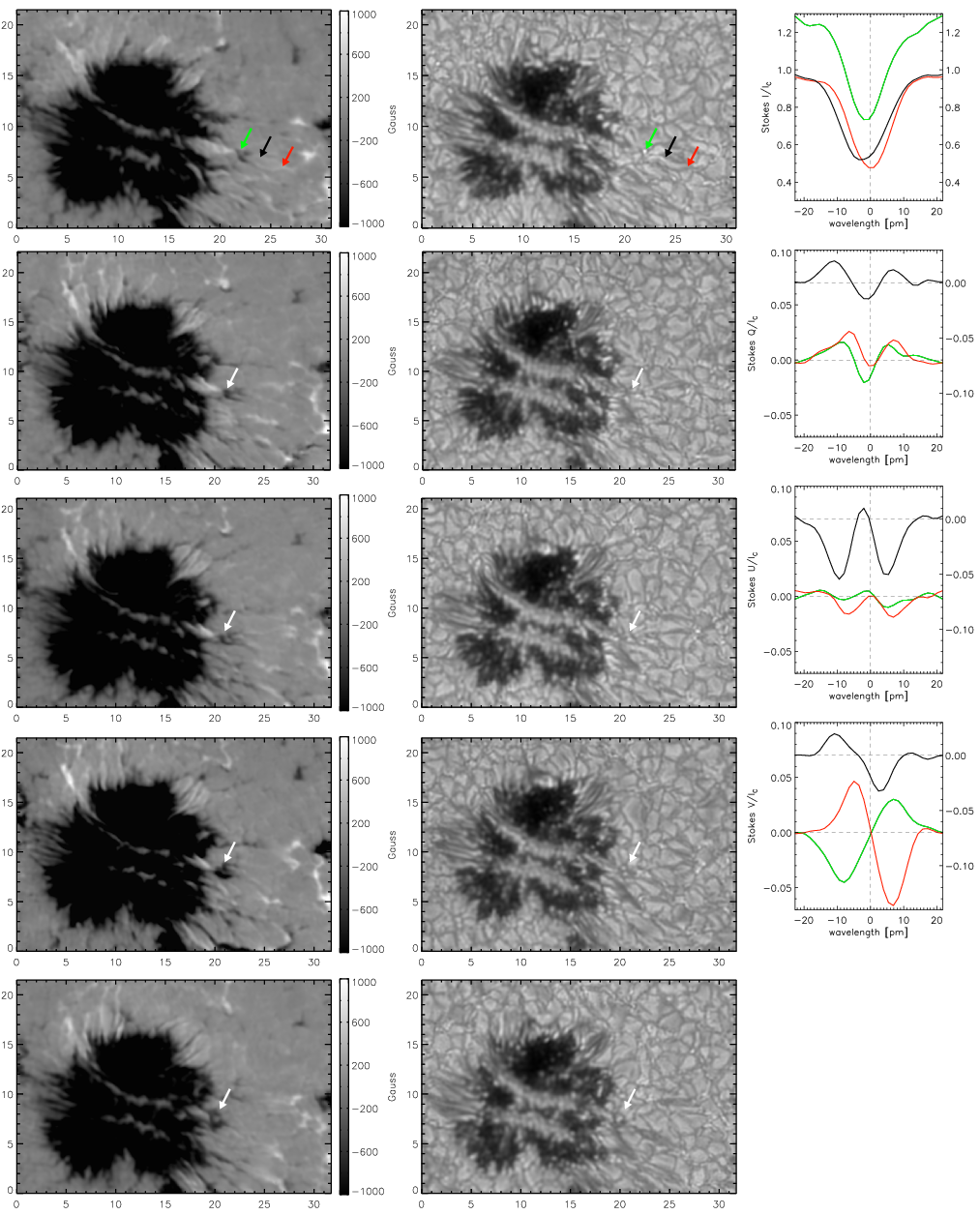}

\caption{Snapshots of the magnetogram and speckle reconstructed continuum image are shown in the left and middle column for 08:51, 09:05, 09:12, 09:15, and 09:30 UT, from top to bottom. The white arrow tracks a black polarity patch as it migrates towards the spot. Tick marks are in arcsec.
Three sets of Stokes profiles ($I(\lambda)$, $Q(\lambda)$, $U(\lambda)$, \& $V(\lambda)$) that correspond to the green, black, and red arrows in the uppermost row of Fig.~\ref{fig:2} are shown in the right column. The green and red lines correspond to the inner and outer end of the elongated granule, respectively, the black line is from the mid-granule. The black $Q$, $U$, and $V$ profiles are shifted by 0.07 (right $y$-axis) for better visibility.
}
\label{fig:2}
\end{figure*}

In a 4:40 hour time series, we observed the formation of a penumbra in the leading spot of NOAA 11024. In Fig.~1 three snapshots of the sunspot evolution are depicted. The penumbra grows in segments continuously, and encircles a little more than half of the umbra at the end of our time series. In \citet{schlichenmaier+al2010a} we demonstrate that the area of the spot increases by 130 arcsec$^2$ from 230 arcsec$^2$. The area increase is exclusively due to an area increase of the penumbra. The area of the umbra including its light bridges is constant in time.

The small pores visible in the images show the same polarity as the spot and mark the direction towards the opposite polarity of the bipolar region NOAA 11024. The penumbra forms on the side away from the active region. In the lower right area of the field-of-view, elongated granules appear and disappear continuously. This is a region of continuous flux emergence. 

\section{Results}

We study the magnetograms from the region of flux emergence, concentrating on elongated granules. We have found a few cases, in which elongated granules are associated with magnetic bipoles.
They are oriented along the line connecting the polarities of the active region, with the pole closer to the spot showing the spot polarity. 
In the case presented here, this bipole separates and most interestingly the spot polarity of the bipole migrates towards the spot.

In the left and middle column of Fig.~\ref{fig:2} we present a 40 min sequence of the evolution displaying the maps of the longitudinal magnetic field (cf. Sect.~\ref{sec:obs}) and speckle-reconstructed broad band images, respectively, at 08:51, 09:05, 09:12, 09:15, and 09:30 UT, from top to bottom. In the magnetogram the gray scale reflects estimated field strength and polarity, clipped at $\pm$ 1000 Gauss.

We track the evolution of one "granular" bipole. The two images at 08:51 in the first row contain three arrows  (with green, black, and red colors)\footnote{In a black and white version of the paper red should correspond to dark gray and green to light gray, respectively.}, which mark an elongated granule. The green arrow marks the closer end of the granule (with respect to a center in the spot), being co-spatial with a bright point in the inter-granular vortex. The red arrow is at the far end of the granule. The closer and far ends of the granule, initially separated by 5 arcsec, are associated with a concentration of magnetic flux of opposite sign as it is seen in the magnetogram.

A set of Stokes profiles corresponding to the elongated granule is shown in the right column of Fig.~\ref{fig:2}. The green and red footpoints show strong $V$ signals of opposite polarity and relatively weak $Q$ and $U$ signals. In the granule, the black profile exhibits strong linear polarization ($Q$ and $U$), and a smaller signal in $V$, connecting the footpoints. This provides evidence that the elongated granule is associated with a loop of magnetic field lines.

The subsequent rows of Fig.~\ref{fig:2} show that the footpoints of the loop separate. The white arrows in the magnetograms at 09:05, 09:12, 09:15, and 09:30 UT show the migration of the black polarity footpoint towards the spot. The white-polarity footpoint moves towards the opposite polarity of the active region and dissolves.
The black-polarity footpoint reaches the spot at the outer edge of the upper light bridge at 09:30 UT. At the outer end of the light bridge, the magnetogram shows opposite polarity initially, but there is no $V$ signal at 09:30.

\section{Discussion} 

We report on flux emergence in the close vicinity of a spot. The proto-spot evolves into a sunspot with a well developed penumbra. We argue that the increase of the magnetic flux of the proto-spot is due to the magnetic flux that emerges through elongated granules in the immediate vicinity of the spot. We present one example in which an elongated granule is associated with the signature of a rising magnetic flux loop. While one foot-point of the loop migrates towards the spot, the other moves in the opposite direction. From this we infer that the small-scale flux emergence contributes to the growth of the sunspot.

Zwaan (1992) reviews the observations of the sunspot formation. He shows that a sunspot forms by coalescence of pores. Here we present an alternative way of flux accumulation to form a sunspot. The small-scale bipolar loops that appear in the emergence site can significantly contribute to the total magnetic flux of the sunspot. Within 4:40 h of observation the spot area grows from 230 arcsec$^2$ to 360 arcsec$^2$, but we do not see a pore that merges with the sunspot. Instead, we witness elongated granules that carry magnetic flux in form of small-scale loops. The distance between the footpoints of some of the small-scale bipoles increases gradually. Finally, the footpoint that has the same polarity as the spot merges with it.

From this we envisage that the increase in the area and magnetic flux of the sunspot during our time sequence is due to merging of footpoints from small-scale bipoles: the footpoints with the proper polarity migrate towards the spot, the footpoints with the other polarity migrate towards the other polarity of the active region. In our case this process contributes one third of the total flux of the sunspot. We have no information on how the pre-existing proto-spot was formed. It may well be that it was formed by merging pores. Hence, both processes are probably relevant for the formation of a fully-fledged sunspot.

We find many examples of elongated granules and rising bipoles, for which the magnetic foot-point closer to spot has the polarity of the spot. 
This seems to indicate that the small-scale emergence at the surface is rooted in a larger coherent structure beneath the surface, as seen in the simulations by \citet{cheung+etal2008}. Mysteriously, all these small-scale bipoles are reassembled at the surface to form a sunspot. Further investigation will help to understand the associated processes of reassembling the magnetic flux into a large coherent structure.

It is also interesting to note that while the magnetic flux merges with the proto-spot on one side, the penumbra forms on the opposite side of the spot. How does the opposite side `know' that flux is accumulated? How does the spot propagate the signal of the incoming magnetic flux to its other end? These questions will be addressed in a further investigation of our spectropolarimetric data set.

\acknowledgements
The German VTT is operated by the Kiepen\-heuer-Institut f\"ur Sonnenphysik at the Spanish Observatorio del Teide. We acknowledge the support by the VTT group, Manolo Collados for setting up TIP, Christian Beck for advice and software, and Oliver Wiloth for assisting the observations. NBG acknowleges the Pakt f\"ur Forschung, and RR the DFG grant Schm 1168/8-2.

%\bibliographystyle{aa} % style aa.bst 
%\bibliography{\bibpath{bibarchive_dec2009.bib}} % your references Yourfile.bib
%\input{AN_noaa11024.bbl}

\end{document}